\documentclass[conference]{IEEEtran}
\usepackage[font=footnotesize]{caption}
\usepackage{url}
\usepackage{verbatim}
\usepackage{subcaption}
\usepackage{soul} 

\usepackage{tikz}
\usetikzlibrary{shapes.geometric, arrows}

\tikzstyle{operator} = [rectangle, rounded corners, minimum width=3cm, minimum height=1cm,text centered, draw=black, fill=blue!30]
\tikzstyle{uav} = [rectangle, rounded corners, minimum width=2cm, minimum height=1cm,text centered, draw=black, fill=green!30]
\tikzstyle{arrow} = [thick,->,>=stealth]

\usepackage{multirow}
\usepackage{array}
\usepackage{mathtools}
\usepackage{mathtools, cuted}

\usepackage{newtxtext} 
\usepackage{newtxmath} 

\usepackage{amsmath}

\usepackage{amsfonts}
\usepackage{amssymb}

\usepackage{bbm}
 \usepackage{stfloats}

\usepackage{amsthm}
\usepackage{array,booktabs}
\usepackage{float}
\usepackage{times}
\usepackage{array}
\usepackage{cite}
\usepackage{dsfont}
\usepackage{multicol}
\usepackage{mathtools, cuted}
\usepackage{float}
\usepackage{breqn}
\usepackage{graphicx}
\usepackage{epstopdf}
\usepackage{mathrsfs}
\usepackage{algorithm}
\usepackage{algorithmic}

\usepackage[english]{babel}
\usepackage[autostyle]{csquotes}
\usepackage{enumerate}

\IEEEoverridecommandlockouts

\renewcommand{\algorithmiccomment}[1]{\bgroup\hfill\tiny//~#1\egroup}
\DeclareMathAlphabet\mathbfcal{OMS}{cmsy}{b}{n}

\usepackage{soul,xcolor}

\usepackage{xcolor}
\usepackage{soul}

\usepackage{tikz}
\usepackage{textcomp}
\usepackage[bookmarks=false]{hyperref}
\usepackage{lipsum}

\begin{document}
\setstcolor{red}
\title{\fontsize{22}{24}\selectfont
AoI-Driven Hierarchical Learning for Cooperative Resource Sharing in Multi-Operator UAV Networks\vspace{-0.8cm}}
\author{\IEEEauthorblockN{
\normalsize Atefeh Hajijamali Arani\IEEEauthorrefmark{1},
 Mahyar Shirvanimoghaddam\IEEEauthorrefmark{2}, 
Abolfazl Mehbodniya\IEEEauthorrefmark{3}, Halim Yanikomeroglu\IEEEauthorrefmark{4},
and
Fumiyuki Adachi\IEEEauthorrefmark{5}}
\IEEEauthorblockA{\small\IEEEauthorrefmark{1}Department of Statistics and Actuarial Science, University of Waterloo, ON N2L 3G1, Canada}
\IEEEauthorblockA{\small\IEEEauthorrefmark{2}School of Electrical and Computer Engineering, The University of Sydney, NSW 2006, Australia}
\IEEEauthorblockA{\small\IEEEauthorrefmark{3}
Department of Electronics and Communication Engineering, Kuwait College of Science and Technology (KCST), Doha, Kuwait
}
\IEEEauthorblockA{\small\IEEEauthorrefmark{4}Carleton-NTN Lab, Department of 
Systems and Computer Engineering, Carleton University, Ottawa, ON K1S 
5B6, Canada}
\IEEEauthorblockA{\small\IEEEauthorrefmark{5}International Research Institute of Disaster Science, Tohoku University,
Sendai, Miyagi, Japan
}
\thanks{Accepted for publication in IEEE GLOBECOM 2026.}
}
\IEEEaftertitletext{\vspace{-1.2\baselineskip}}

\maketitle 

\begin{abstract}
Uncrewed aerial vehicle (UAV)-assisted networks provide a versatile paradigm for on-demand connectivity. However, in multi-operator aerial networks (MOANs), the joint optimization of cooperative resource sharing and 3D trajectory control to maintain information freshness is a complex combinatorial problem, which can be shown to be NP-hard. To address this computational complexity, we propose an age of information (AoI)-driven hierarchical deep reinforcement learning (DRL) framework. Specifically, a Dueling Double Deep Q-Network (D3QN) architecture is deployed at both the operator and UAV decision layers to mitigate overestimation bias and enhance stability in high-dimensional state spaces. To improve system resilience, we introduce an AoI- and load-aware outage compensation mechanism that prioritizes users based on instantaneous transmission demands and temporal freshness. Furthermore, a normalized load exchange balance metric is incorporated to regulate cooperative behavior and ensure resource fairness across operators. 
Simulation results demonstrate that the proposed hierarchical D3QN significantly outperforms conventional DRL, non-cooperative, and cooperative benchmarks, reducing the average AoI by up to 56.1\% under severe congestion while ensuring superior inter-operator fairness and outage mitigation.
\end{abstract}
\begin{IEEEkeywords}
Age of information, cooperative resource sharing, deep reinforcement learning, multi-operator aerial networks. %
\vspace{-0.5cm}
\end{IEEEkeywords}

\section{Introduction} \vspace{-0.1cm}
{Uncrewed aerial vehicles (UAVs) have emerged as a flexible platform for enhancing wireless connectivity, owing to their rapid deployment, adaptive three-dimensional (3D) mobility, and on-demand coverage provisioning\cite{uav_Maritime2026,drl_llm_2026}. In applications such as environmental monitoring, public safety, and industrial Internet of Things (IoT), UAVs can operate as aerial base stations (BSs) to complement terrestrial infrastructure and improve service continuity \cite{uav_Non-Ideal_Backhaul2026}. Nevertheless, next-generation services such as real-time sensing, mission-critical control, and digital twin operation require not only high data rates but also timely and \emph {fresh} information at the receiver. In this context, conventional metrics (e.g., throughput) do not explicitly capture information timeliness. Age of information (AoI) has therefore attracted increasing attention as a destination-centric metric 
quantifying the elapsed time since the generation of the 
most recently received status update \cite{Marco12_infocom}.
 Although UAV-assisted wireless networks have been widely investigated, their operation in \emph{multi-operator} environments introduces substantial technical challenges. 
 In multi-operator aerial networks (MOANs), multiple service providers manage their own UAVs and coordinate resource sharing to improve coverage and reliability, particularly during user roaming, traffic overload, or emergency deployments.
However, inter-operator coordination introduces strategic and operational coupling. User association and load distribution across multiple UAVs directly affect scheduling feasibility and, consequently, AoI performance. Persistent load imbalance may lead to overload conditions, user dropping, and delayed status updates, thereby increasing the AoI experienced by affected users. Furthermore, in the absence of explicit fairness control, resource sharing may result in \emph{cross-subsidization}, where one operator allocates service capacity to another at the expense of its own users' information freshness \cite{zhang2024age}. \looseness=-1}
 {Despite these challenges, most existing studies on UAV trajectory optimization and resource allocation primarily consider single-operator settings and optimize sum-rate maximization \cite{Rui18_tcom}.  Although learning-based methods have been adopted for 3D trajectory optimization and throughput enhancement, they typically do not explicitly integrate AoI considerations into the control framework in multi-operator environments \cite{Huaiyu18_infocom}. 
 Furthermore, the joint user association, scheduling, and trajectory design in MOANs is a mixed-integer, non-convex problem where centralized solutions are computationally intractable due to coupled variables and limited scalability with network size  \cite{Gesbert18_spawc}. Motivated by these limitations, we propose an AoI-driven hierarchical deep reinforcement learning (DRL) framework for  resource sharing and UAV trajectory optimization in MOANs. The proposed architecture operates over two time scales: I) at a coarse time-window level, operators learn long-term resource-sharing policies; II) at a fine time instant level, UAVs learn trajectory actions to serve users. To ensure balanced cooperation, we incorporate a mechanism that controls cross-operator load exchange under limited inter-operator information sharing.
The key technical contributions are summarized as follows:\looseness=-1}
\begin{itemize}
\setlength{\itemsep}{0pt}
\setlength{\parskip}{0pt}
\item \textit{AoI-aware MOAN framework:} We develop a multi-operator UAV model explicitly capturing AoI dynamics under strict load constraints and outage compensation.
\item \textit{Hierarchical DRL design:} We propose a two-tier Dueling Double Deep Q-Network (D3QN) framework to optimize inter-operator resource sharing and UAV 3D trajectories.
\item \textit{AoI- and load-aware outage compensation:} We design a novel dropping policy prioritizing both transmission demand and information staleness to mitigate excessive AoI growth during network overload.
\item \textit{Load exchange balance (LEB) metric:} We introduce a regulatory metric to penalize one-sided load transfer and prevent cross-subsidization
between cooperating operators.
\item \textit{Privacy-preserving coupling:} Inter-operator resource sharing is executed using only aggregate data to preserve operator-level privacy. 
\end{itemize}
The remainder of this paper is organized as follows. 
Section \ref{sys_model_section} presents the system model and outage mechanism. Section \ref{prob_formulation_sec} formulates the optimization problem, solved using the D3QN framework developed in Section \ref{DRL_sec}. Section \ref{sec_results} discusses the simulation results, and Section \ref{sec_conc} concludes the paper.

\vspace{-0.2cm}
 \section{System Model} \label{sys_model_section}\vspace{-0.1cm}
We consider the downlink transmission phase of a multi-operator aerial network (MOAN). The system comprises two independent operators, indexed by $\mathcal{Q}=\{1,2\}$, each deploying a set of UAV-mounted  BSs to provide wireless connectivity and timely status updates. Let $\mathcal{U}_q$ denote the set of UAVs owned by operator $q \in \mathcal{Q}$, with the set of total UAV defined as $\mathcal{U}=\bigcup_{q\in\mathcal{Q}} \mathcal{U}_q$. The UAVs fly at a fixed speed $v_{u}$ and dynamically adjust their 3D trajectories to accommodate spatial traffic fluctuations and maintain information freshness. 
Let $\mathcal{K}$ denote the set of all users and $\mathcal{L}_q$ represent the subset of users subscribed to operator $q \in \mathcal{Q}$, such that $\mathcal{K} = \bigcup_{q\in\mathcal{Q}} \mathcal{L}_q$ and $\mathcal{L}_q \cap \mathcal{L}_{q'} = \emptyset$, $\forall q \neq q'$. While each user is subscribed to a single operator, they may be served by any UAV in $\mathcal{U}$ depending on the active inter-operator resource sharing policy. We assume that the time horizon is discretized into consecutive windows indexed by $j$ with duration $T_w$, each comprising $N_w$ time instants indexed by $t$ with duration $T_s = T_w/N_w$. This two-timescale structure accommodates the different dynamics of the network, where inter-operator resource sharing occurs at the coarse window level and UAV mobility is managed at the fine time instant level.\vspace{-0.1cm}
\subsection{Radio Propagation and Signal Quality}
\vspace{-0.1cm}
Let $\mathbf{z}_u(t) = (x_u(t), y_u(t), h_u(t))$ and $\mathbf{z}_k(t) = (x_k(t), y_k(t), h_k)$ denote the 3D coordinates of UAV $u$ and user $k$ at time $t$, respectively. The air-to-ground  channel is modeled using a probabilistic line-of-sight (LoS)/non-line-of-sight (NLoS) framework to account for environmental blockages.
The horizontal distance between UAV $u$ and user $k$ at time $t$ is given by $  r_{u,k}(t) = \sqrt{(x_u(t)-x_k(t))^2 + (y_u(t)-y_k(t))^2},$
and the corresponding 3D distance is $d_{u,k}(t) = \sqrt{r_{u,k}^2(t) + (h_u(t)-h_k)^2}.$
The probability of establishing a LoS link is expressed as \cite{ITU_model} \vspace{-0cm}
\begin{equation}\label{prob_LoS}
\mathrm{pr}_{u,k}^{(\mathrm{LoS})}(t)
=
\prod_{n=0}^{J}
\left[
1-\exp\left(
-\frac{\left[h_u(t)-\frac{(n+\frac{1}{2})(h_u(t)-h_k)}{J+1}\right]^2}{2\xi^2}
\right)
\right],
\end{equation}
where $J = \left\lfloor \frac{r_{u,k}(t)\sqrt{\alpha\beta}}{1000} - 1 \right\rfloor$
approximates the expected number of blocking obstacles along the horizontal path between UAV $u$ and user $k$ at time $t$. The environmental parameters $\alpha$, $\beta$, and $\xi$ characterize the propagation environment, representing the building area fraction, the average building density, and the standard deviation of building heights, respectively. 
The large-scale channel gain between UAV $u$ and user $k$, under propagation condition $z \in \{\mathrm{LoS}, \mathrm{NLoS}\}$, is modeled in dB as \cite{azari_2019} \vspace{-0cm}
\begin{equation}\label{channel_access}
L_{u,k}^{(z)}(t)=\delta^{(z)}+\eta^{(z)}\log_{10}d_{u,k}(t)+\chi_{u,k}^{(z)}(t),  \vspace{-0.1cm}
\end{equation}
where $\delta^{(z)}$ and $\eta^{(z)}$ denote the path-loss constant and exponent, respectively. The small-scale fading $\chi_{u,k}^{(z)}(t)$ (in dB) assumes a Nakagami-$m$ fading model, where the fading power follows a Gamma distribution with severity parameter $m^{(z)}$, and is assumed to be independent and identically distributed (i.i.d.) across consecutive time instants. 
The corresponding linear channel gain is given by $g_{u,k}(t) = 10^{-L_{u,k}^{(z)}(t)/10}$. Each UAV  $u$  is equipped with a single omnidirectional antenna and transmits with a fixed power $p_u$ over an allocated bandwidth $\omega$. To enable structured multi-operator coexistence and eliminate inter-operator interference, orthogonal spectrum allocation across operators is assumed. Consequently, co-channel interference originates only from UAVs belonging to the same operator that reuse the spectrum.
Under this assumption, the received signal-to-interference-plus-noise ratio (SINR) at user $k$, when served by UAV $u \in \mathcal{U}$ at time $t$, is expressed as \vspace{-0.1cm}
\begin{equation}
\gamma_{u,k}(t)
=
\frac{p_u g_{u,k}(t)}
{\sum\limits_{\substack{u' \in \mathcal{U}\backslash u}} 
p_{u'} g_{u',k}(t)\rho_{u'}(t)
+ \sigma^2}, \vspace{-0.1cm}
\end{equation}
where $\rho_{u'}(t)$ denotes the load of the interfering UAV $u'$ at time $t$, and $\sigma^2$ represents the noise power.
The achievable rate of user $k$ served by UAV $u$ is then given by the Shannon capacity formula \vspace{-0cm}
\begin{equation}
C_{u,k}(t)
=
\omega \log_2 \!\left(1 + \gamma_{u,k}(t)\right). \vspace{-0.1cm}
\end{equation}
 \subsection{Load and User Association Modeling}\vspace{-0cm}
Each user $k$ is associated with a traffic demand characterized by an average arrival rate $\lambda_k$ (packets/s) and mean packet size $F_k$ (bits), The corresponding average data rate requirement is defined as $\Lambda_k = \lambda_k F_k$.
Given the instantaneous achievable rate $C_{u,k}(t)$, the fraction of transmission time required by UAV $u$ to serve user $k$ at time $t$ is expressed as \vspace{-0.2cm}
\begin{equation} \label{user_density_eq}
\delta_{u,k}(t) = \frac{\Lambda_k}{C_{u,k}(t)}. \vspace{-0.2cm}
\end{equation}
Let $\mathcal{K}_u(t)$ denote the set of users associated with UAV $u$ at time $t$. Thus, the aggregate load of UAV $u$ is given by \vspace{-0cm}
\begin{equation}
\rho_u(t) = \sum_{k \in \mathcal{K}_u(t)} \delta_{u,k}(t). \vspace{-0cm}
\end{equation}
The load $\rho_u(t)$ represents the normalized resource utilization of UAV $u$ within the considered time interval. To ensure feasible scheduling, the load must satisfy $0 \le \rho_u(t) \le 1$.
Since the achievable rate $C_{u,k}(t)$ depends on interference from neighboring UAVs, which is itself a function of their respective loads, the network load vector $\boldsymbol{\rho}(t) = \big(\rho_1(t), \dots, \rho_{|\mathcal{U}|}(t)\big)^T$ is inherently coupled. This non-linear system is solved via a standard fixed-point iteration, $\boldsymbol{\rho}^{(\iota)} = \min\!\left(\boldsymbol{f}\big(\boldsymbol{\rho}^{(\iota-1)}\big), \mathbf{1}\right)$, where $\mathbf{1}$ denotes the all-ones vector and $\iota$ denotes the iteration index. Under standard interference mapping properties, this guarantees convergence to a unique feasible load satisfying $\rho_u(t) \le 1, \forall u \in \mathcal{U}$\cite{atefeh_tvt2026}.
 
User association directly affects the load distribution and, consequently, the interference coupling across UAVs. At each time $t$, user $k$ selects a serving UAV from the available set $\mathcal{U}^*$, which is determined by the inter-operator sharing policy. Under restricted cross-operator access, $\mathcal{U}^*$ is limited to the UAVs of the operator to which the user belongs; otherwise, $\mathcal{U}^* = \mathcal{U}$.
To jointly account for link quality and load conditions, the serving UAV is selected according to \cite{Sumudu16} \vspace{-0cm}
\begin{equation}
u_k^*(t) = \arg\max_{u \in \mathcal{U}^*}
\left\{
p_u g_{u,k}(t)\big(1 - \hat{\rho}_u(t)\big)
\right\}, \vspace{-0cm}
\end{equation}
where $\hat{\rho}_u(t)$ denotes the estimated load of UAV $u$ at time $t$. The load estimate is updated recursively as \vspace{-0.15cm}
\begin{equation}
\hat{\rho}_u(t)
=
\hat{\rho}_u(t-1)
+
\nu(t)\big(\rho_u(t-1)-\hat{\rho}_u(t-1)\big), \vspace{-0cm}
\end{equation}
where $\nu(t)$ is the learning rate for load estimation. 
When inter-operator resource sharing is enabled, users subscribed to operator $q$ may be served by UAVs of another operator $q'$. To quantify this resulting inter-operator load transfer, we define the load exchanged during window $j$ as 
\vspace{-0cm}
\begin{equation}
\rho_{q \rightarrow q'}(j)
=
\sum_{t \in \mathcal{T}_j}
\sum_{\substack{k \in \mathcal{L}_q \\ u \in \mathcal{U}_{q'}}}
\delta_{u,k}(t),\vspace{-0.2 cm}
\end{equation}
where $\mathcal{T}_j$ denotes the set of time instants within window $j$.
Similarly, the load received by operator $q$ from operator $q'$ during window $j$ is denoted by $\rho_{q' \rightarrow q}(j)$.
To balance the load exchange between the operators, we define the load exchange balance (LEB) metric. The LEB metric for operator $q$ in time window $j$ can be expressed as  \vspace{-0.15cm}
\begin{equation}
\mathrm{LEB}_q(j)
=
\left|
\rho_{q' \rightarrow q}(j)
-
\rho_{q \rightarrow q'}(j)
\right|. \vspace{-0.1cm}
\end{equation}
A smaller value of LEB reflects more balanced inter-operator cooperation. Nevertheless, the absolute value function is non-differentiable at zero, which may adversely affect gradient-based optimization. To overcome this limitation, a smoothed approximation of the LEB metric is introduced as follows:
\vspace{-0cm}
\begin{equation}
\widetilde{\mathrm{LEB}}_q(j)
=
\sqrt{
\left(
\rho_{q' \rightarrow q}(j)
-
\rho_{q \rightarrow q'}(j)
\right)^2
+
\epsilon}, \vspace{-0.15cm}
\end{equation}
where $\epsilon > 0$ is a small constant ensuring differentiability and numerical stability.
Given that the load imbalance is lower bounded by zero and upper bounded by $\max_{q\in\mathcal Q} |\mathcal U_q| T_w$, the metric is normalized to satisfy $0 \le \widehat{\widetilde{\mathrm{LEB}}}_q(j) \le 1$ as  $\widehat{\widetilde{\mathrm{LEB}}}_q(j)
=
\frac{
\widetilde{\mathrm{LEB}}_q(j)
-
\sqrt{\epsilon}
}{
\sqrt{\left(\max_{q\in\mathcal Q} |\mathcal U_q| T_w\right)^2+\epsilon}
-
\sqrt{\epsilon}
}$\footnote{Note that while the load exchange metrics and the overall system model are formulated for a two-operator scenario ($\mathcal{Q}=\{1,2\}$) to maintain notational tractability, the proposed framework can be readily extended to an arbitrary number of operators ($|\mathcal{Q}| > 2$) by aggregating the inter-operator resource exchanges.\looseness=-1}. \vspace{-0.0cm}
\subsection{Age of Information and Outage Compensation} \vspace{-0.05cm}
To quantify information freshness from the perspective of the users, we utilize the AoI metric. For each user $k\in \mathcal K$, the instantaneous AoI at time $t$, denoted by $\Delta_k(t)$, is defined as the time elapsed since the generation of the most recently received status update. Let $\mathbbm{1}_k(t) \in \{0,1\}$ be the service indicator, where $\mathbbm{1}_k(t)=1$ if user $k$ is successfully scheduled without experiencing a channel-induced or congestion-based drop, and $0$ otherwise. To reflect practical staleness limits and maintain analytical tractability, we impose a maximum threshold $\Delta_{\max}$. The discrete-time AoI evolution is thus expressed as \vspace{-0cm}
\begin{equation} \label{aoi_eq}
\Delta_k(t+1)=
\begin{cases}
1, & \text{if } \mathbbm{1}_k(t)=1,\\
\min\big(\Delta_k(t)+1,\, \Delta_{\max}\big), & \text{otherwise}.
\end{cases} \vspace{-0.15cm}
\end{equation}
 While AoI quantifies timeliness, it is fundamentally constrained by network capacity. When a UAV becomes overloaded, i.e., $\rho_u(t) > 1$, an outage compensation mechanism must be adopted to enforce the feasibility constraint. Conventional schemes typically drop users based solely on their transmission demand in \eqref{user_density_eq}, which may lead to excessive AoI growth for time-sensitive users \cite{Sumudu16,atefeh_globecom18}. To overcome this limitation, we adopt an AoI- and load-aware dropping policy that jointly considers the fractional transmission requirement and the  AoI of users. 
 For each user $k \in \mathcal{K}_u(t)$, a priority metric is defined as \vspace{-0cm}
\begin{equation}
\pi_k(t) = \kappa_1 \delta_{u,k}(t) - \kappa_2 \frac{\Delta_k(t)}{\Delta_{\max}}, \vspace{-0cm}
\end{equation}
where ${\kappa}_1, \kappa_2 \in [0, 1]$ are weighting coefficients such that $\kappa_1 + \kappa_2 = 1$. Users are ranked according to $\pi_k(t)$ and iteratively dropped until the capacity constraint $\rho_u(t) \le 1$ is satisfied. By penalizing the dropping metric for users with stale information, this mechanism provides a tunable trade-off between load regulation and AoI preservation.
 \vspace{-0cm}
\section{Problem Formulation} \label{prob_formulation_sec} \vspace{-0cm}
The proposed MOAN aims to minimize user AoI while maintaining inter-operator fairness. Due to dynamic topologies, interference coupling, and cross-operator load exchange, the resulting optimization problem is highly non-convex and computationally intractable for real-time centralized solvers. Therefore, we cast the problem as a two-timescale hierarchical optimization framework: the upper layer optimizes resource-sharing policies at the coarse window level, while the lower layer performs real-time UAV trajectory and scheduling control at the fine time-instant level.\vspace{-0cm}
\subsection{Operator-Level Optimization}\vspace{-0cm}
At the beginning of each time window $j$, operator $q \in \mathcal{Q}$ selects its inter-operator access policy. The action space is discrete and defined as $\mathcal{A}_q = \{a_q^{\text{self}}, a_q^{\text{cross}}\}$, where $a_q^{\text{self}}$ restricts users to be served exclusively by UAVs of operator $q$, and $a_q^{\text{cross}}$ allows users to access UAVs of other operators.
To ensure fairness among scheduled users, Jain's Fairness Index is adopted. Let $\overline{C}_{k,q}(j) = \frac{1}{N_w} \sum_{t \in \mathcal{T}_j} \sum_{u \in \mathcal{U}_q} C_{u,k}(t)$ denote the average achievable rate of user $k$ served by operator $q$ during window $j$. The rate fairness index is given by \vspace{-0cm}

\begin{equation}
J_q(j) =
\frac{\left( \sum_{k \in \mathcal{K}_q(j)} \overline{C}_{k,q}(j) \right)^2}
{|\mathcal{K}_q(j)| \sum_{k \in \mathcal{K}_q(j)} \left( \overline{C}_{k,q}(j) \right)^2},
\end{equation}
where $\mathcal{K}_q(j)$ denotes the set of users served by operator $q$ during window $j$.
Since rate fairness alone does not capture timeliness requirements, the average AoI is incorporated into the operator objective. 
Let $\overline{\Delta}_q(j) = \frac{1}{|\mathcal{K}_q(j)| N_w} \sum_{t \in \mathcal{T}_j} \sum_{k \in \mathcal{K}_q(j)} \Delta_k(t)$ denote the mean AoI of users served by operator $q$ over window $j$. Given the bounded AoI model with threshold $\Delta_{\max}$, a normalized freshness term is defined as $1 - \frac{\overline{\Delta}_q(j)}{\Delta_{\max}}$, which maps the AoI metric into the interval $[0,1]$.
To jointly capture throughput fairness, AoI, and cooperative balance, the operator utility at window $j$ is formulated as\vspace{-0cm}
\begin{equation}\label{eq:operator_utility}
U_q(j) =
w_1 J_q(j)
+
w_2  (1 - \frac{\overline{\Delta}_q(j)}{\Delta_{\max}})
+
w_3 (1 - \widehat{\widetilde{\mathrm{LEB}}}_q(j) ),\vspace{-0.1cm}
\end{equation}
where $w_1, w_2, w_3\!\in\![0, 1]\!$ are weighting factors balancing fairness, AoI, and load exchange, respectively, satisfying $\sum_{i=1}^{3}\!\! w_i =\!1$. \vspace{-0.45cm}
\subsection{UAV-Level Optimization}\vspace{-0.1cm}
At the UAV level, each UAV dynamically adjusts its 3D trajectory to maintain load feasibility and minimize local AoI. For each UAV $u \in \mathcal{U}$, the discrete action space $\mathcal{A}_u$ comprises seven possible motion decisions, defined as \vspace{-0cm}
\begin{equation} \label{action_uav_eq}
\mathcal{A}_u = \{\mathrm{forward}, \mathrm{backward}, \mathrm{left}, \mathrm{right}, \mathrm{up}, \mathrm{down}, \mathrm{hover}\}. \vspace{-0.2cm}
\end{equation}
The UAV trajectory directly influences the channel gains and, consequently, both the UAV load and the users' AoI. To quantify this timeliness, let $\overline{\Delta}_u(t) = \frac{1}{|\mathcal{K}_u(t)|} \sum_{k \in \mathcal{K}_u(t)} \Delta_k(t)$ denote the mean AoI of users associated with UAV $u$ at time $t$. To jointly capture the load of the UAV and information freshness, the  utility of UAV $u$  is formulated as \vspace{-0cm}
\begin{equation}\label{eq:uav_utility}
U_u(t) =
\mu_1(1 - \rho_u(t))
+
\mu_2 (1 - \frac{\overline{\Delta}_u(t)}{\Delta_{\max}}),\vspace{-0.15cm}
\end{equation}
where $\mu_1, \mu_2\!\!\in\![0, 1]$ are weighting coefficients satisfying $\mu_1 + \mu_2\!=\!1$. The first term penalizes excessive load to prevent network congestion, while the second term promotes local AoI reduction.
\section{Deep Learning-Based Hierarchical Optimization} \label{DRL_sec} \vspace{-0cm}
To realize the hierarchical two-timescale framework described in Section \ref{prob_formulation_sec}, we develop a model-free multi-agent DRL architecture. The proposed framework employs the  D3QN algorithm at both the operator and UAV levels to learn optimal control policies over their respective decision horizons. At the upper layer, each operator acts as an independent D3QN agent, selecting its inter-operator access strategy at the beginning of each coarse time window based on the observed network states and performance indicators. At the lower layer, each UAV acts as a local agent, determining its trajectory at each fine time instant to adapt to dynamic channel conditions, load variations, and AoI evolution. Through continuous interaction with the environment, the agents iteratively update their networks to learn policies that jointly optimize resource sharing and trajectory control within the hierarchical structure.
\vspace{-0cm}
\subsection{Markov Decision Process  Formulation}\vspace{-0.1cm}
To apply DRL, the decision-making processes for both the operators and the UAVs must be formulated as distinct, yet coupled, Markov Decision Processes (MDPs), defined by their respective state spaces, action spaces, and reward functions.
\subsubsection{Operator-Level MDP}
At the window level, each operator $q \in \mathcal{Q}$ is modeled as an agent making decisions at discrete time windows indexed by $j$. The operator-level decision process is defined as follows:
\begin{itemize}
\item \textit{State space} $\mathcal{S}_q$:  
The state at window $j$ captures load conditions, inter-operator load exchange, the LEB metric, and information freshness, and is defined as \vspace{-0.1cm}
\begin{equation}\resizebox{0.8\hsize}{!}{$
s_q(j) =
\left\{
\hat{\rho}_{q}(j),
\hat{\rho}_{q \rightarrow q'}(j),
\hat{\rho}_{q' \rightarrow q}(j),
\widehat{\widetilde{\mathrm{LEB}}}_q(j),
\hat{\overline{\Delta}}_q(j)
\right\}$}.\vspace{-0.1cm}
\end{equation}
Here, $\hat{\rho}_{q}(j)$ denotes the normalized internal load of operator $q$ generated by its own users within window $j$. The terms $\hat{\rho}_{q \rightarrow q'}(j)$ and $\hat{\rho}_{q' \rightarrow q}(j)$ represent the normalized transmitted and received inter-operator loads, respectively. The quantity $\widehat{\widetilde{\mathrm{LEB}}}_q(j)$ denotes the normalized smoothed LEB, and $\hat{\overline{\Delta}}_q(j) = \overline{\Delta}_q(j)/\Delta_{\max}$ represents the normalized average AoI of users  served by operator $q$.

\item \textit{Action space} $\mathcal{A}_q$:  
The action $a_q(j) \in \{a_q^{\text{self}}, a_q^{\text{cross}}\}$ determines the association policy for the subsequent window, specifying whether users are restricted to UAVs of operator $q$ (i.e., $a_q^{\text{self}}$) or allowed cross-operator access (i.e., $a_q^{\text{cross}}$).

\item \textit{Reward function} $\mathcal{R}_q$:  
The immediate reward $r_q(j)$ is defined by the operator utility in \eqref{eq:operator_utility}, yielding a bounded scalar feedback that jointly reflects rate fairness, information freshness, and cooperative balance.
\end{itemize}
\subsubsection{UAV-Level MDP}
At the UAV level, each UAV $u\!\in\!\mathcal{U}$ is modeled as an  agent operating at discrete time instants indexed by $t$. The UAV-level decision process is defined as follows:

\begin{itemize}

\item \textit{State space} $\mathcal{S}_u$:  
The state at time $t$ captures spatial position, load condition, and AoI, and is defined as \vspace{-0cm}
\begin{equation}
s_u(t) =
\left\{
\hat{x}_u(t),
\hat{y}_u(t),
\hat{h}_u(t),
{\rho}_u(t),
\overline{\Delta}_u(t)/\Delta_{\max}
\right\}. \vspace{-0cm}
\end{equation}
Here, $\hat{x}_u(t)$, $\hat{y}_u(t)$, and $\hat{h}_u(t)$ denote the normalized 3D coordinates of UAV $u$.
\item \textit{Action space} $\mathcal{A}_u$:  
The action $a_u(t)$ corresponds to a discrete 3D motion step defined in \eqref{action_uav_eq}.
\item \textit{Reward function} $\mathcal{R}_u$:  
The instantaneous reward $r_u(t)$ is defined by the local utility function in \eqref{eq:uav_utility}, yielding a scalar feedback that jointly evaluates load feasibility and information freshness.
\end{itemize}\vspace{-0cm}
\subsection{Hierarchical Dueling Double Deep Q-Network Design}\vspace{-0cm}
Classical tabular Q-learning becomes computationally intractable in the considered MOAN framework due to the continuous high-dimensional state spaces. 
Standard Deep Q-Networks (DQNs) suffer from severe overestimation bias, creating a divergent feedback loop in highly dynamic aerial environments. To ensure stable convergence, we deploy a hierarchical D3QN architecture for operator and UAV agents, which explicitly decouples state-value and action-advantage estimation to mitigate this bias.
Specifically, the state $s$ is first processed through shared fully connected feature extraction layers parameterized by $\boldsymbol{\theta}$. The network is then decomposed into two parallel streams: a state-value stream $V(s; \boldsymbol{\theta}, \beta)$ and an advantage stream $A(s, a; \boldsymbol{\theta}, \alpha)$, where $\beta$ and $\alpha$ represent their respective trainable parameters. The final Q-value is reconstructed using an aggregation layer \cite{Cavdar2026}: \vspace{-0.2cm}
\begin{equation}
Q(\!s,\!a;\!\boldsymbol{\theta}, \alpha,\!\beta)\!\!=\!\!V(\!s; \boldsymbol{\theta},\! \beta)\!+\!A(\!s,\!a;\!\boldsymbol{\theta}, \alpha)-\frac{1}{|\mathcal{A}|}\!\!\sum_{a'\!\in\!\mathcal{A}}\!\!\!A(s, a'\!;\!\boldsymbol{\theta}, \alpha\!).\vspace{-0cm}
\end{equation}
During training, mini-batches of transition tuples $e=(s,a,r,s')$ are sampled from an experience replay buffer $\mathcal{D}$. To overcome overestimation bias, the double Q-learning mechanism decouples action selection from target evaluation. The target $y$ is computed as 
$y=r+\gamma Q(s^{\prime},\arg\max_{a^{\prime}}Q(s^{\prime},a^{\prime};\theta,\alpha,\beta);\theta^{-},\alpha^{-},\beta^{-}),$
where $\gamma\!\!\in\!(0,1)$ is the discount factor, and $\boldsymbol{\theta}^-$ denotes the parameters of the target network. The policy network parameters are optimized via gradient descent by minimizing the loss between the predicted Q-values and the target $y$. Rather than executing periodic hard replacements, the target network parameters are continuously updated using Polyak averaging as $\boldsymbol{\theta}^-\!\leftarrow\!\tau \boldsymbol{\theta} + (1 - \tau) \boldsymbol{\theta}^-$, where $\tau\!\ll\!1$. Finally, to balance exploration and exploitation, the agents utilize an $\epsilon$-greedy action selection strategy with an exponentially decaying exploration probability.

\subsection{Hierarchical Coupling of Operator and UAV Layers}
Although operating over different timescales, the operator and UAV decision processes are inherently coupled. At the operator tier, the window-level action $a_q(j)$ dictates the inter-operator association policy, directly conditioning the instantaneous load and local AoI observed by the UAVs. Conversely, the UAVs' sequential trajectory actions $\{a_u(t)\}$ determine the time-varying channel gains, rates, and load realizations. These frame-level outcomes accumulate to construct the operator-level reward metrics, including fairness, AoI, and load exchange balance. Crucially, resource sharing relies solely on load exchange aggregate data to preserve operator-level privacy. 
\vspace{-0cm}
\section{Performance Evaluation} \label{sec_results} \vspace{-0cm}
We evaluate the proposed AoI-driven hierarchical D3QN framework in a $500 \times 500$ m$^2$ MOAN with uniformly distributed users. Unless otherwise specified, the network defaults to 2 UAVs per operator. The simulation parameters and network configurations are summarized in Table~\ref{tab:sim_params}. 
In the proposed approach, a shared 128-neuron fully connected layer branches into parallel 64-neuron state-value and advantage streams,  using layer normalization, ReLU activation, and a 0.2 dropout rate.
The networks are trained via RMSprop (learning rate $0.001$) with a Smooth L1 loss function, while target networks are continuously updated with a soft update coefficient of $\tau = 0.005$.
Performance is compared against five baselines: 
1) \textit{DDQN}: Utilizes the proposed hierarchical structure and rewards, but employs standard Double Deep Q-Network (DDQN) agents;
2) \textit{Load-Aware DQN}: Employs standard DQN agents and utilizes an outage compensation mechanism based solely on fractional transmission requirement;
3) \textit{AoI-Aware Non-Cooperative 3D-UAV}: Restricts operators to serving only their subscribed users ($a_q(j) = a_q^{\text{self}}$), while UAVs employ DQN agents for 3D trajectory optimization alongside the proposed AoI-aware outage compensation policy;
4) \textit{Load-Aware Non-Cooperative 3D-UAV}:
Identical to baseline 3, but employs strictly load-based outage compensation;
5) \textit{Load-Aware Cooperative 2D-UAV}: Enforces full cooperation ($a_q(j) = a_q^{\text{cross}}$) for global UAV access, combining DQN-optimized 2D fixed-altitude ($h_{\max}$) trajectories with load-based outage compensation.  Results are averaged over multiple Monte Carlo runs using an Intel Xeon Gold 6230R CPU (512 GB RAM) and an NVIDIA RTX 3090 GPU (24 GB VRAM).\looseness=-1\vspace{-0.1cm}

\begin{table}[tb!]
\vspace{0.1cm}
\caption{Simulation Parameters}\vspace{-0cm}
\label{tab:sim_params}
\centering
\resizebox{\columnwidth}{!}{%
\begin{tabular}{ll l l}
\toprule
\textbf{Parameter} & \textbf{Value} & \textbf{Parameter} & \textbf{Value} \\
\midrule
Network area & $500 \times 500$ m$^2$ & Path-loss ($\delta^{\mathrm{NLoS}}, \eta^{\mathrm{NLoS}}$) & $32.9, 37.5$ \\
Number of operators & $2$ & Nakagami-$m$ fading ($m^{\mathrm{LoS}}, m^{\mathrm{NLoS}}$) & $3, 1$ \\
UAV altitude, speed & $22.5{\sim}121.9$m,  $10$m/s & Maximum AoI ($\Delta_{\max}$),  $N_w$  & $50, 5$  \\
Transmission power, bandwidth & $24$ dBm, $10$ MHz &   $\kappa_1,  \kappa_2$ & $0.3, 0.7$ \\
Noise power spectral density & $-174$ dBm/Hz & $w_1, w_2, w_3$& $1/3, 1/3, 1/3$\\
Path-loss ($\delta^{\mathrm{LoS}}, \eta^{\mathrm{LoS}}$) & $41.1, 20.9$ & D3QN $\gamma$, batch size & 0.99, 128 \\
\bottomrule
\end{tabular}%
}
\vspace{-0.45cm}
\end{table}
\vspace{-0cm}\subsection{Results Analysis} 
\begin{figure}[tb!]
    \centering
    \includegraphics[width=\columnwidth]{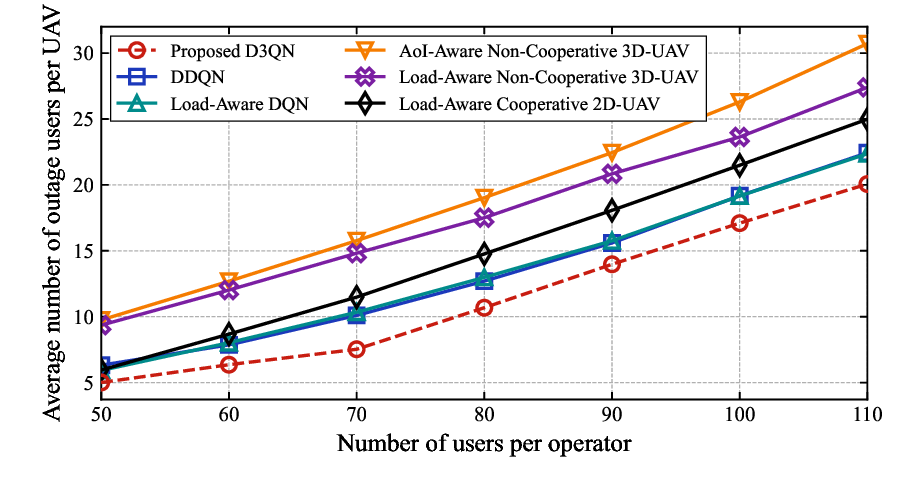}\vspace{-0.5cm}
    \caption{Average outage per UAV versus the number of users per operator for a network with 2 UAVs per operator.}    \label{fig:outage_ue}\vspace{-0cm}
\end{figure}
Fig. \ref{fig:outage_ue} illustrates the average number of users experiencing outages per UAV as a function of the number of users per operator. 
Here, the average outage is defined as the mean number of users dropped by UAVs to satisfy the load feasibility constraint $\rho_u(t) \le 1, \forall u \in \mathcal{U}$. We observe that the outage increases with user density for all schemes due to the limited resources and higher load levels on UAVs. However, the proposed hierarchical D3QN framework consistently achieves the lowest outage across all network scales, demonstrating its superior capability in jointly optimizing resource sharing and UAV trajectory. Specifically, the performance gap becomes more pronounced as the number of users increases, highlighting the scalability of the proposed approach under congested conditions. Compared to DDQN and Load-Aware DQN, the proposed D3QN effectively mitigates overestimation bias and leverages AoI-aware decision-making, resulting in more efficient load distribution and reduced user outages. In contrast, non-cooperative schemes exhibit significantly higher outage levels due to the lack of inter-operator resource sharing.  Furthermore, the Load-Aware Cooperative 2D-UAV scheme, while benefiting from cooperation, suffers from limited trajectory flexibility due to its fixed-altitude constraint, resulting in inferior performance compared to the proposed 3D optimization framework.
\begin{figure}[tb!]
\vspace{0.1cm}
    \centering
    \includegraphics[width=\columnwidth]{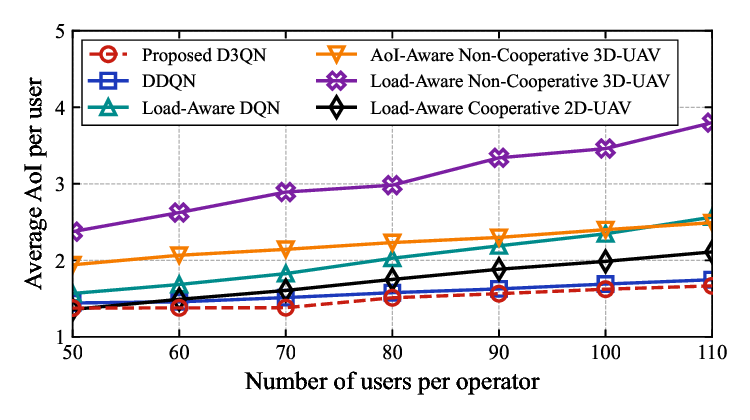}\vspace{-0.5cm}
    \caption{Average AoI per user versus the number of users per operator for a network with 2 UAVs per operator.}
    \label{fig:aoi_ue} 
\end{figure}

Fig. \ref{fig:aoi_ue} shows the average per-user AoI versus user density. Although increased user densities raise AoI in all schemes, the proposed D3QN consistently maintains the lowest AoI. Specifically, under severe congestion (110 users per operator), D3QN reduces the average AoI by 56.1\% and 33.1\% compared to the load-aware and AoI-aware non-cooperative baselines, respectively, demonstrating the necessity of inter-operator sharing. Furthermore, D3QN outperforms the load-aware DQN by 35\%, the load-aware cooperative 2D-UAV by 21\%, and the hierarchical DDQN by 4.6\%. These significant gains confirm that combining 3D trajectory flexibility, AoI-aware outage compensation, and D3QN's stable value estimation effectively mitigates service delays.
\begin{figure}[tb!]
    \centering
    \includegraphics[width=\columnwidth]{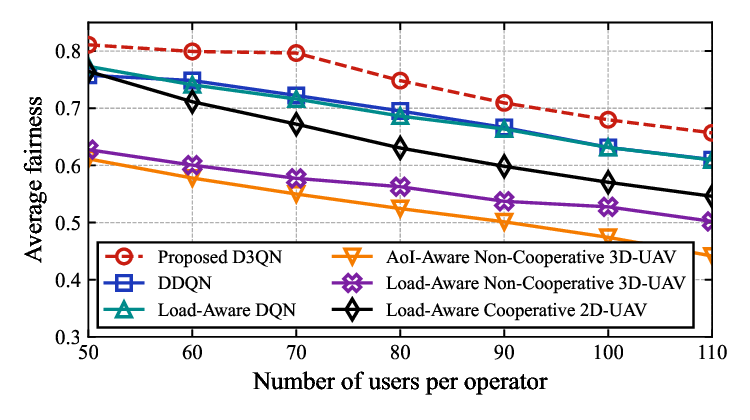}\vspace{-0.5cm}
    \caption{Average fairness versus the number of users per operator for a network with 2 UAVs per operator.}
    \label{fig:fairness_ue} 
\end{figure}

Fig. \ref{fig:fairness_ue} plots the average fairness versus user density. While resource competition degrades fairness in denser networks, the proposed D3QN consistently maintains the most balanced allocation. Specifically, at 70 users per operator, D3QN improves fairness by 44.9\% and 38.0\% over the AoI-aware and load-aware non-cooperative schemes, respectively, which exhibit the lowest fairness due to the absence of load sharing. Moreover, it yields gains of 18.5\% over the load-aware cooperative 2D-UAV which is limited by reduced trajectory flexibility as well as 11.2\% over the load-aware DQN, and 10.3\% over the hierarchical DDQN.

\begin{figure}[tb!]
    \centering
    \includegraphics[width=\columnwidth]{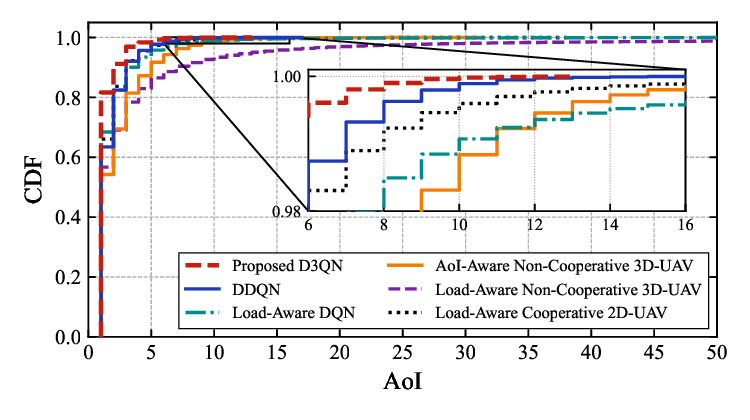}\vspace{-0.5cm}
    \caption{CDF of the per-user AoI for a network with 2 UAVs and 70 users per operator.}    \label{fig:AoI_CDF} 
\end{figure}

Fig. \ref{fig:AoI_CDF} presents the CDF of the AoI for all schemes. The proposed D3QN curve is consistently left-shifted compared to all baselines, indicating a higher probability of achieving lower AoI values and thus superior information freshness.
The improvement is particularly evident in the tail region (highlighted in the inset), where the proposed method significantly reduces the occurrence of high AoI values, demonstrating enhanced reliability and robustness. In contrast, non-cooperative schemes exhibit heavier tails, reflecting a higher likelihood of stale information due to load imbalance and lack of resource sharing. Although DDQN and load-aware DQN improve performance relative to non-cooperative approaches, they remain inferior to the proposed method due to less accurate value estimation and the absence of fully integrated AoI-aware control.

\begin{figure}[tb!]
    \centering
    \includegraphics[width=\columnwidth]{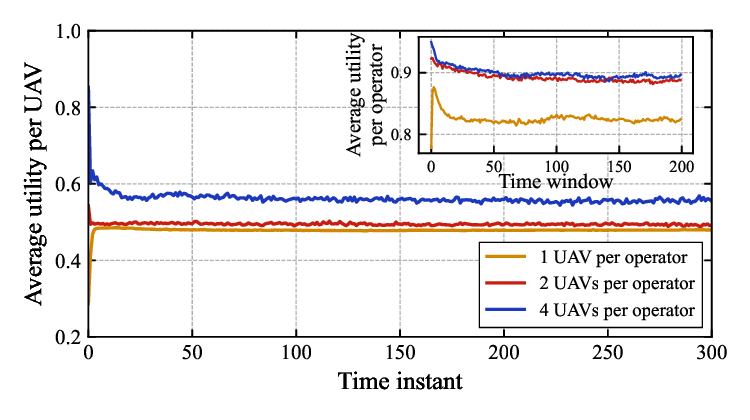} \vspace{-9mm}
    \caption{Convergence of the average utility per UAV (main) and the average utility per operator (inset) under varying UAV deployment densities for the proposed D3QN approach with 50 users per operator.}
    \label{fig:conv50} 
\end{figure}

Fig.~ \ref{fig:conv50}  illustrates the convergence behavior of the proposed hierarchical D3QN framework in terms of both UAV-level and operator-level utilities under different network densities (i.e., 1, 2, and 4 UAVs per operator) with 50 users per operator. The main plot shows the average utility per UAV over time instants. It is observed that UAV agents rapidly converge to stable trajectories and scheduling policies within the first 30 time steps across all scenarios. Moreover, the average UAV utility increases with the number of deployed UAVs due to enhanced spatial flexibility and improved load balancing. The inset presents the average utility per operator over time windows, where operator-level agents also exhibit stable convergence. Deployments with 2 or 4 UAVs achieve significantly higher utilities compared to the single-UAV case. 
\section{Conclusion} \label{sec_conc}  
This paper proposes an AoI-driven hierarchical D3QN framework for cooperative resource sharing and 3D UAV trajectory optimization in MOANs. Furthermore, a novel AoI- and load-aware outage compensation mechanism is developed to maintain information freshness under strict capacity constraints. By integrating AoI-aware control, fairness, and load balancing, the proposed approach effectively addresses the coupled challenges of AoI and resource sharing. Simulation results demonstrate significant improvements over learning-based, non-cooperative, and cooperative baselines in terms of outage, AoI, and fairness, particularly in dense network conditions.

\bibliographystyle{IEEEtran}
\bibliography{references.bib}

\begin{thebibliography}{10}
\providecommand{\url}[1]{#1}
\csname url@samestyle\endcsname
\providecommand{\newblock}{\relax}
\providecommand{\bibinfo}[2]{#2}
\providecommand{\BIBentrySTDinterwordspacing}{\spaceskip=0pt\relax}
\providecommand{\BIBentryALTinterwordstretchfactor}{4}
\providecommand{\BIBentryALTinterwordspacing}{\spaceskip=\fontdimen2\font plus
\BIBentryALTinterwordstretchfactor\fontdimen3\font minus \fontdimen4\font\relax}
\providecommand{\BIBforeignlanguage}[2]{{%
\expandafter\ifx\csname l@#1\endcsname\relax
\typeout{** WARNING: IEEEtran.bst: No hyphenation pattern has been}%
\typeout{** loaded for the language `#1'. Using the pattern for}%
\typeout{** the default language instead.}%
\else
\language=\csname l@#1\endcsname
\fi
#2}}
\providecommand{\BIBdecl}{\relax}
\BIBdecl

\bibitem{uav_Maritime2026}
H.~Pan, B.~Lin, J.~An, and G.~Sun, ``Joint deployment, association and power optimization for {UAV}-mounted {STAR-RIS}-assisted maritime communications,'' \emph{IEEE Trans. Veh. Technol. (Early Access)}, pp. 1--16, 2026.

\bibitem{drl_llm_2026}
Y.~Gong, J.~Fan, R.~Zhang, D.~Niyato, Y.~Yao, and X.~Chang, ``Safe and economical {UAV} trajectory planning in low-altitude airspace: A hybrid {DRL-LLM} algorithm with compliance awareness,'' \emph{IEEE Trans. Mob. Comput. (Early Access)}, pp. 1--17, 2026.

\bibitem{uav_Non-Ideal_Backhaul2026}
Y.~Jiang, N.~Deng, H.~Wei, N.~Zhao, and A.~Nallanathan, ``A drl approach to multi-{UAV} deployment for cellular systems with non-ideal backhaul,'' \emph{IEEE Commun. Lett.}, vol.~30, pp. 1195--1199, 2026.

\bibitem{Marco12_infocom}
S.~Kaul, R.~Yates, and M.~Gruteser, ``Real-time status: How often should one update?'' in \emph{Proc. IEEE INFOCOM}, 2012, pp. 2731--2735.

\bibitem{zhang2024age}
M.~Zhang, H.~H. Yang, A.~Arafa, and H.~V. Poor, ``Age of information in mobile networks: Fundamental limits and tradeoffs,'' in \emph{MobiHoc '24}, 2024, pp. 321--330.

\bibitem{Rui18_tcom}
Q.~Wu and R.~Zhang, ``Common throughput maximization in {UAV}-enabled {OFDMA} systems with delay consideration,'' \emph{IEEE Trans. Commun.}, vol.~66, no.~12, pp. 6614--6627, 2018.

\bibitem{Huaiyu18_infocom}
J.~Liu, X.~Wang, B.~Bai, and H.~Dai, ``Age-optimal trajectory planning for {UAV}-assisted data collection,'' in \emph{IEEE INFOCOM WKSHPS}, 2018, pp. 553--558.

\bibitem{Gesbert18_spawc}
H.~Bayerlein, P.~De~Kerret, and D.~Gesbert, ``Trajectory optimization for autonomous flying base station via reinforcement learning,'' in \emph{IEEE SPAWC}, 2018, pp. 1--5.

\bibitem{ITU_model}
{ ITU-R P.1410-5}, ``Propagation data and prediction methods required for the design of terrestrial broadband radio access systems operating in a frequency range from 3 to 60 {GHz},'' Feb. 2012.

\bibitem{azari_2019}
M.~M. Azari, F.~Rosas, and S.~Pollin, ``Cellular connectivity for {UAVs}: Network modeling, performance analysis, and design guidelines,'' \emph{IEEE Trans. Wirel. Commun.}, vol.~18, no.~7, pp. 3366--3381, 2019.

\bibitem{atefeh_tvt2026}
A.~H. Arani, X.~Fernando, O.~Alhussein, and Y.~Zhu, ``Deep reinforcement learning for resource sharing and {UAV} trajectory optimization in multi-operator {UAV}-assisted wireless networks,'' \emph{IEEE Trans. Veh. Technol. (Early Access)}, pp. 1--16, 2026.

\bibitem{Sumudu16}
S.~Samarakoon, M.~Bennis, W.~Saad, and M.~Latva-aho, ``Dynamic clustering and on/off strategies for wireless small cell networks,'' \emph{IEEE Trans. Wirel. Commun.}, vol.~15, no.~3, pp. 2164--2178, 2016.

\bibitem{atefeh_globecom18}
A.~H. Arani, A.~Mehbodniya, M.~J. Omidi, and M.~F. Flanagan, ``Satisfaction based channel allocation scheme for self-organization in heterogeneous networks,'' in \emph{IEEE GLOBECOM}, 2018, pp. 1--6.

\bibitem{Cavdar2026}
Y.~Deng, S.~Zhang, I.~A. Meer, M.~Ozger, and C.~Cavdar, ``Joint trajectory and handover management for {UAVs} co-existing with terrestrial users: A multi-agent {DRL} approach,'' \emph{IEEE T. Cogn. Commun. Netw.}, vol.~12, pp. 1195--1210, 2026.

\end{thebibliography}

\end{document}